\documentclass[10pt,conference]{IEEEtran}
\usepackage{svg}
\usepackage{amssymb}
\usepackage[utf8]{inputenc}
\usepackage[english]{babel}
\usepackage{cite}
\usepackage{amsmath,amssymb,amsfonts}
\usepackage{subcaption}
\usepackage{graphicx}
\usepackage{textcomp}
\usepackage{xcolor}
\usepackage{mathrsfs}
\usepackage[utf8]{inputenc}
\usepackage{comment}
\usepackage[ruled,vlined]{algorithm2e}
\usepackage{mathtools}
\usepackage[font=small,labelfont=bf]{caption}
\usepackage{blkarray, bigstrut}
\usepackage{cases}
\usepackage{amsmath}

\title{Cooperative Jamming for Physical Layer Security Enhancement Using Deep Reinforcement Learning}

 \author{\IEEEauthorblockN{Sayed Amir Hoseini$^{1}$, Faycal Bouhafs$^{1}$, Neda Aboutorab$^{2}$, Parastoo Sadeghi$^{2}$ and Frank den Hartog$^{1}$}\\
 \IEEEauthorblockA{ $^{1}$School of Systems and Computing \\
 $^{2}$School of Engineering and Technology \\
University of New South Wales, Canberra, Australia \\
 Email: \{s.a.hoseini,  f.bouhafs, n.aboutorab, p.sadeghi, frank.den.hartog\}@unsw.edu.au}}

\begin{document}
\bstctlcite{IEEEexample:BSTcontrol}
\maketitle

\begin{abstract}
Wireless data communications are always facing the risk of eavesdropping and interception. Conventional protection solutions which are based on encryption may not always be practical as is the case for wireless IoT networks or may soon become ineffective against quantum computers. In this regard, Physical Layer Security (PLS) presents a promising approach to secure wireless communications through the exploitation of the physical properties of the wireless channel. Cooperative Friendly Jamming (CFJ) is among the PLS techniques that have received attention in recent years. However, finding an optimal transmit power allocation that results in the highest secrecy is a complex problem that becomes more difficult to address as the size of the wireless network increases. In this paper, we propose an optimization approach to achieve CFJ  in large Wi-Fi networks by using a Reinforcement Learning Algorithm. Obtained results show that our optimization approach offers better secrecy results and becomes more effective as the network size and the density of Wi-Fi access points increase.
\end{abstract}

\section{Introduction}
The pervasive and broadcasting nature of electromagnetic waves in wireless networks poses a significant risk of unauthorized access and interception of sensitive data. Although encryption and authentication are common security measures, Physical Layer Security (PLS) is increasingly being recognized as a promising approach to provide an additional layer of protection to wireless communications~\cite{liu2016physical}. However, until recently PLS was primarily restricted to theoretical study of the problem.

We have shown in~\cite{ hoseini2022ccnc,hoseini2023demo} that PLS could indeed be realized by exploiting the flexibility offered by Software-Defined Networking (SDN) and Spectrum Programming~\cite{bouhafs2018wi}. In our previous work, we have shown that it is possible to implement a PLS solution for Wi-Fi networks using off-the-shelf equipment~\cite{hoseini2022ccnc,hoseini2023demo}. In these works, we implemented an algorithm that associates the legitimate station to the Access Point (AP) that provides the highest secrecy capacity in the presence of an eavesdropper~\cite{Faycal2020globecom}. 
We extended these works in~\cite{hoseini2022jamming} by incorporating the idea of cooperative friendly jamming (CFJ), where APs produce jamming signals to further degrade the capacity of eavesdropping.  However, this  work was limited to a single user, a single eavesdropper, and two APs. The optimization formulation and closed-form equations presented in~\cite{hoseini2022jamming} are not scalable to more realistic scenarios that include multiple APs, legitimate stations, and eavesdroppers.

In this paper, we extend our contribution in~\cite{hoseini2022jamming} to larger wireless networks, where APs can work as both legitimate traffic sources and jammers at the same time. More specifically, we aim to optimize the transmit power of each AP that results in the maximum achievable sum secrecy capacity across legitimate users. For this purpose, we propose to use a Reinforcement Learning (RL) technique within Machine Learning (ML) to find a near-optimal radio configuration. In our proposed RL framework, the state of the network is the location of nodes, which underpins the received power at each legitimate or eavesdropper node according to the assume wireless propagation path-loss model. The action taken is the transmit power from each AP and the revenue is the sum secrecy capacity among legitimate users. Numerical evaluation of our proposed RL method in numerous network scenarios shows significant sum secrecy capacity improvement across the network.  

The rest of the paper is structured as follows. In Section~\ref{sec:ralatedWork}, the related work is presented briefly.  Section~\ref{sec:model} formulates the proposed CFJ, optimization problem, and the reinforcement learning model, followed by simulation results in Section~\ref{sec:result}. Finally, we conclude the paper and discuss the future work in Section~\ref{sec:conc}.

\section{Related Work}
\label{sec:ralatedWork}
The application of ML to wireless communications has gained significant interest over the last few years~\cite{RN1}. Several ML-based solutions have been proposed in the literature to help address issues such as spectrum sensing~\cite{RN2}, RF coexistence~\cite{RN3}, and RF transmitter identification~\cite{RN4}. 
In the context of jamming, contributions focused only on proposing ML-based solutions to detect jamming attacks and their perpetrators~\cite{RN5,RN6,RN7}.  
Authors in~\cite{RN8} proposed to identify possible ML techniques that could be applied to PLS, but focused on Physical Layer Authentication, Antenna Selection, and Relay Node Selection.  In~\cite{RN9}, authors proposed an ML-based PLS to secure communications between Intelligent Reflecting Surfaces and Mobile Edge Computing nodes. In~\cite{wang2022secrecy}, authors proposed a federated learning framework to enhance PLS where part of the model aims to optimize the transmit power to reduce the eavesdropping opportunity.

In addition, all contributions mentioned above were limited to theoretical models. Our works in~\cite{hoseini2022ccnc,hoseini2023demo} were the first implementation of PLS in wireless networks. In these works, we presented an implementation of a PLS solution where a legitimate station is associated with the AP that offers the highest secrecy in the presence of an eavesdropper. The choice of the AP is based on the algorithm we proposed in~\cite{Faycal2020globecom} and the implementation exploits the spectrum programmability concept presented in~\cite{bouhafs2018wi}.  In~\cite{hoseini2022jamming},  we extended our previous work by proposing a friendly jamming method that improves the secrecy of wireless communication in the presence of an eavesdropper. In~\cite{hoseini2022jamming}, we used an optimization model that determines the transmit power that a jamming AP should use to improve the secrecy capacity of the connection. However, this model could only be applied to a scenario that involves two APs and one legitimate user and could not scale to larger deployment scenarios.  In this current study, we improve upon this model by proposing a more comprehensive wireless network solution that can enhance PLS through CFJ. We employ Reinforcement Learning~\cite{haarnoja2018SAC} to determine the best power allocation to APs. Most existing works on power optimization, such as~\cite{eletreby2015supporting,adams2017FJ}, considered a limited point-to-point scenario (i.e., Alice-Bob communication). To the best of our knowledge, there is no previous work on joint optimization of user association and power allocation for a scenario with multiple APs and jammers. Therefore, the results in our work aim to show the superiority of the proposed model to existing systems in the real world or those proposed in the literature. Furthermore, as far as we know, this is the first research to propose and solve this model using reinforcement
learning.

\section{System Model}\label{sec:model}
\subsection{Proposed Friendly Jamming Formulation}

\begin{figure}[h]
    \centering
    % \includesvg[width=0.4\textwidth]{figures/diagram.svg}
    \includegraphics[trim=155 250 50 120, clip, width=0.4\textwidth]{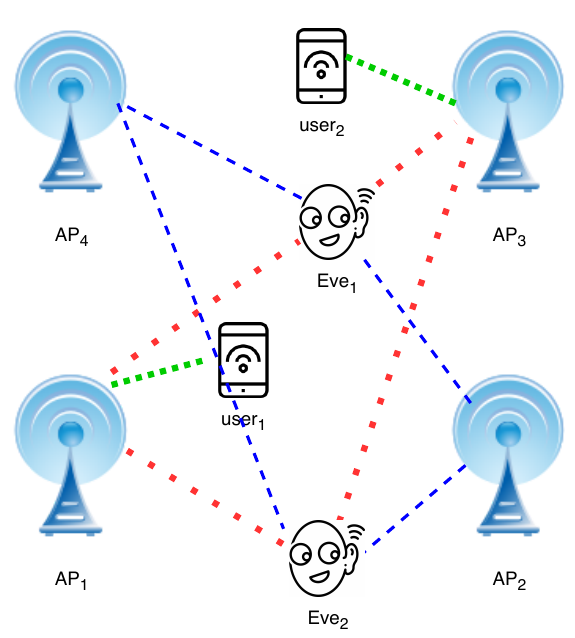}
    \caption{In a cooperative wireless network, user$_1$ and user$_2$ receive downlink traffic from AP$_1$ and AP$_3$, respectively (shown by green dotted lines). However, Eve$_1$ and Eve$_2$ are eavesdropping on their traffic (indicated by red dotted lines). Additionally, AP$_2$ and AP$_4$ are idle and act as jammers. It is important to note that all APs and users communicate in the same frequency band. As a result, AP$_1$'s signal can also be considered as jamming signal for any eavesdropper wiretapping AP$_3$'s traffic, and AP$_2$'s signal can also be considered as jamming signal for any eavesdropper wiretapping AP$_1$'s traffic.}
    \label{fig:diagram}
\end{figure}

Let us assume a network that includes  $N$ APs  (AP$_1$, AP$_2$, $\dotsc$, AP$_N$ ),  $K$ users as $u_1$, $u_2$, $\dotsc$, $u_K$, and $J$ eavesdroppers as $e_1$, $e_2$, $\dotsc$, $e_J$, which are located in the area that is serviced by a cooperative wireless network. Each user $u_k$ is associated with $AP_{\alpha_k}$, where $A=[\alpha_1,\ldots,\alpha_K$] is a vector that denotes the association of users to APs and $P^t=[p^t_1,\ldots, p^t_N]$ is a vector that represents the transmit power of each AP. APs transmit signals continuously. If there is an associated user to receive data they act as a normal AP and otherwise as a jammer. A simple network with $N=4$, $K=2$ and $J=2$ is illustrated in \figurename~\ref{fig:diagram}.

We  consider the Friis transmission formula \cite{rappaport2002mobile} for the wireless propagation model where for a transmit power of $p^t$, the received power at distance $d$ is calculated by 
\begin{equation}
    p^r=p^tG_tG_r(\frac{\lambda}{4\pi})^2(\frac{1}{d})^\gamma,
\end{equation}
where $G_t$ and $G_r$ are antenna gain for the transmitter and receiver respectively, $\lambda$ is the wavelength of the radio frequency and $\gamma$ is the path loss exponent. The received power at $u_k$ and $e_j$ from $AP_n$ are denoted by $p^{ru}_{n,k}$ and $p^{re}_{n,j}$, respectively.  Therefore, the Shannon capacity~\cite{cover2006elements} of the downlink channel between AP$_n$ and the legitimate user $u_k$  is given as 
\begin{equation}\label{eq:legit_cap}
\begin{split}
     C_{n,k} &= W\log(1+\text{SINR}_{n,k})\\ 
     &= W\log\left(1+\frac{p^{ru}_{n,k}}{ \sum^{N}_{\nu=1, \nu\neq n }p^{ru}_{\nu,k}+N_{u_k}}\right),
\end{split}
\end{equation}
where $W$ is the channel bandwidth, SINR$_{n,k}$ is the Signal-to-Interference-plus-Noise Ratio (SINR) at $u_{k}$ from AP$_{n}$ and $N_{u_k}$ is the noise power at $u_{k}$. Similarly, the Shannon capacity of the channel between AP$_n$ and the eavesdropper $e_j$ is
\begin{equation}\label{eq:eve_cap}
\begin{split}
        C^e_{n,j}&=W\log(1+\text{SINR}_{n,j}) \\
        &=W\log\left(1+\frac{p^{re}_{n,j}}{ \sum^{N}_{\nu=1, \nu\neq n }p^{re}_{\nu,j}+N_{e_j}}\right),
\end{split}
\end{equation}
where SINR$_{n,j}$ is the SINR at $e_{j}$ from AP$_{n}$ and $N_{e_j}$ is the noise power at $e_{j}$. In this paper, all $\log$s are in base 2 and, therefore, capacities are measured in bits/s. Equation~\eqref{eq:eve_cap} shows all other APs contribute to the jamming and reduce the eavesdropper capacity, regardless of whether the AP is functioning as a jammer or transmitting data to users. We assume there is no collusion among eavesdroppers. In this paper, we consider the wiretap capacity~\cite{wyner1975wire} as the PLS performance metric, which is widely studied in the literature, e.g., see~\cite{vilela2011wireless} for an early work and~\cite{PLS_VLC_2022} for more recent results.
Also, let us assume the worst-case scenario when the eavesdropper with the highest capacity is wiretapping. The maximum capacity among all eavesdroppers to wiretap the traffic transmitted by $AP_n$ is defined by 
\begin{equation}\label{eq:maxEveCap}
    C^{e}(n)=\max(\{C^e_{n,j}: j = 1,\ldots,J\}).
\end{equation} 
Assuming the user $u_k$ is associated with $AP_{\alpha_k}$, the secrecy capacity for the user $u_k$ is given by
\begin{equation}
    C^{s}(u_k,\alpha_k) = [{C_{\alpha_k,k}-C^{e}(\alpha_k)}]^{+},
\end{equation}
where $[x]^{+}$ denotes $\max\{x,0\}$. 
In this paper, we aim to find the joint optimal user association and power allocation to the APs that  maximize 
the sum of the secrecy capacity of all legitimate users, which the optimization problem is formulated as follows
\begin{equation}    \label{eq:sum_cam_opt}
\begin{split}
    \Psi:\quad &\max_{P^t,A}  (\sum^{K}_{k=1} C^{s}(u_k,\alpha_k)),\\
           & \text{s.t.} \quad 0 \leq p^t_n \leq p_{max},\\
           &\quad \quad \alpha_k \in \{1,2, ...,N\},
\end{split}
\end{equation}
where $A:=[\alpha_1,\ldots,\alpha_K$]. Solving this joint optimization problem is a difficult task due to the non-linear and non-convex nature of secrecy capacity expression~\cite{zhang2013optimization} where the user association optimization adds extra complexity. In this paper, we propose to decouple the original problem into two subproblems. First, we employ the AP selection algorithm from~\cite{Faycal2020globecom,hoseini2022ccnc} where each user $u_k$ is associated with  $AP_{\alpha_k}$ that provides the maximum secrecy capacity $C^{s}(u_k)$ considering eavesdropper locations and a uniform power allocation to APs (i.e., $p^t_n=p_{max}$). That is, the association index $\alpha_k$ for user $u_k$ is obtained by
\begin{equation}
    \label{eq:AP_sel}
    \alpha_k=\arg\max_{n \in \{1,2, ...,N\}}  ({C_{n,k}-C^{e}(n)}).
\end{equation}
Then, we use this user association $A=[\alpha_1,\ldots,\alpha_K$] and formulate the optimization of the power allocation of APs by
\begin{equation}    \label{eq:sum_cam_opt_power}
\begin{split}
    \Phi  =&\arg\max_{P^t=[p^t_1,\ldots, p^t_N]}  (\sum^{K}_{k=1} C^{s}(u_k,\alpha_k)),\\
           & \text{s.t.} \quad 0 \leq p^t_n \leq p_{max},
\end{split}
\end{equation}
which is still a non-convex problem. To solve this, We employ Deep Reinforcement Learning that maximizes the sum secrecy capacity. This approach lets us extend this solution in our future works to consider user mobility or channel selection.

\subsection{Proposed Reinforcement Learning Model}
\label{sec:RL-SAC}
Since the wireless network is controlled by software, APs, and stations' information are available for the controller's software that can be used to control the network globally~\cite{bouhafs2018wi}. Therefore, the reinforcement learning model for our proposed system is defined as follows:
\begin{itemize}
    \item \textbf{State} ($\mathbf{S}$) is defined based on the observed information of the environment that includes the location of users, eavesdroppers, and APs. Thus, we define the state as $\mathbf{S}=\{L_{AP}, L_{u},L_{e}\}$ where $L_{AP}=[l_{AP1},l_{AP2},\ldots, l_{APN}]$ is a vector that denotes the location of APs, $L_u=[l_{u1},l_{u2},\ldots, l_{uK}]$ typifies the location of user stations, and $L_e=[l_{e1},l_{e2},\ldots, l_{eJ}]$ represents the location of eavesdroppers.
    \item \textbf{Action} ($\mathbf{a}$) is defined as the vector of transmit power of APs: $P^t=[p^t_1,\ldots, p^t_N]$.
    \item \textbf{Revenue} ($\mathbf{R}$) is the combination of rewards and penalties after taking action $\mathbf{a}$ at state $\mathbf{S}$. In our model, we only consider a reward for the sum of positive secrecy capacities among users:
    \begin{equation}
    \label{eq:revenue}
        R(\mathbf{S},\mathbf{a}) = \sum^{K}_{k=1} C^{s}(u_k,\alpha_k).
    \end{equation}
\end{itemize}

This paper utilizes the Soft Actor-Critic (SAC) algorithm \cite{haarnoja2018SAC}, which is an off-policy approach resulting in a stochastic policy for optimal actions. The SAC algorithm's main advantage is the ability to balance exploration and exploitation, which the model can adjust during training to prevent the policy from converging too early to an undesirable local optimum \cite{SACopenAI}.

\section{Performance Evaluation}
\label{sec:result}

\begin{figure*}
   \centering
    \begin{subfigure}{0.27\textwidth}
        \includegraphics[trim=0 0 0 0, clip, height=5.6cm]{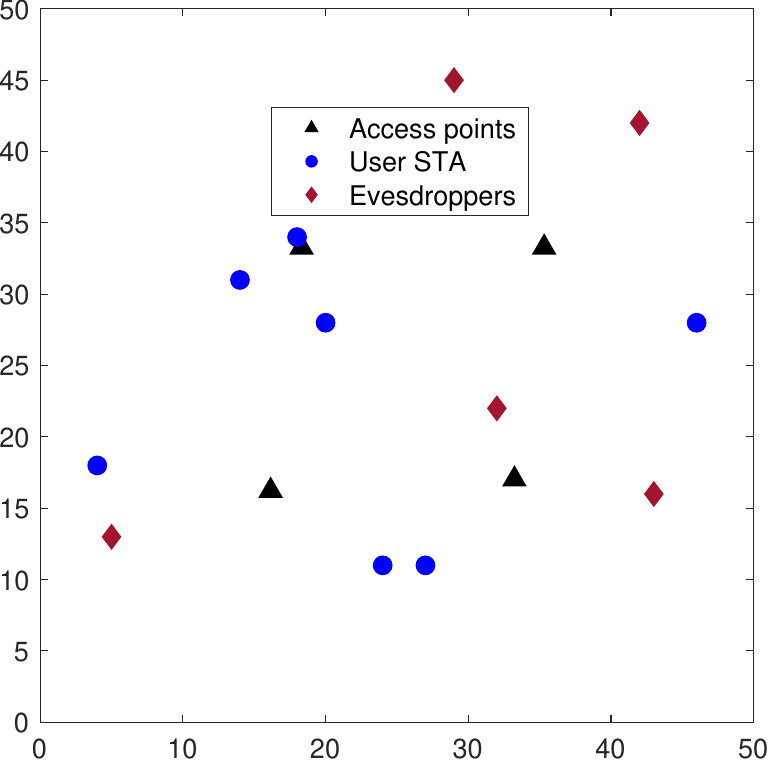}
        \caption{scenario 1: 4 APs.}
    \end{subfigure}
    \hspace{25pt}
        \begin{subfigure}{0.27\textwidth}
        \includegraphics[trim=0 0 0 0, clip, height=5.6cm]{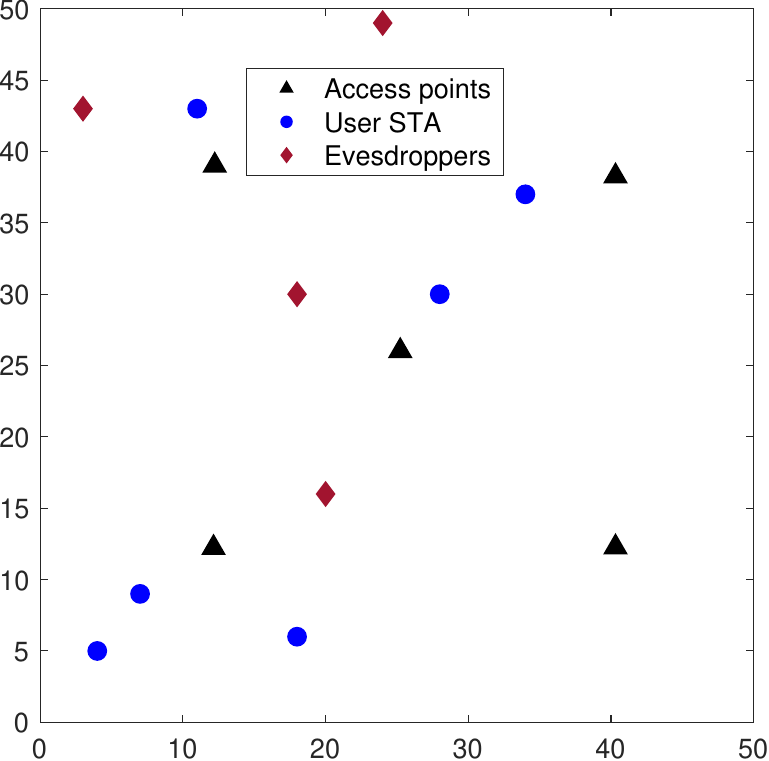}
        \caption{scenario 2: 5 APs.}
    \end{subfigure} 
    \hspace{25pt}
    % \begin{subfigure}{0.3\textwidth}
    %     \includegraphics[trim=0 0 0 0, clip, height=5.6cm]{figures/sch_sc3.pdf}
    %     \caption{scenario 3}
    % \end{subfigure} \\
    \begin{subfigure}{0.27\textwidth}
        \includegraphics[trim=0 0 0 0, clip, height=5.6cm]{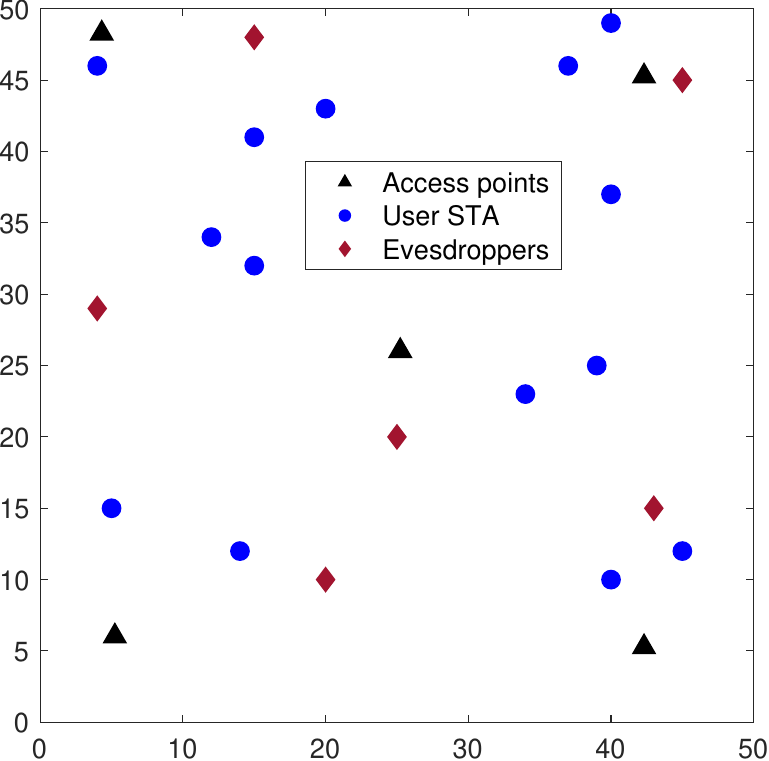}
        \caption{scenario 3: 5 APs.}
    \end{subfigure} \\
    \begin{subfigure}{0.27\textwidth}
        \includegraphics[trim=0 0 0 0, clip, height=5.6cm]{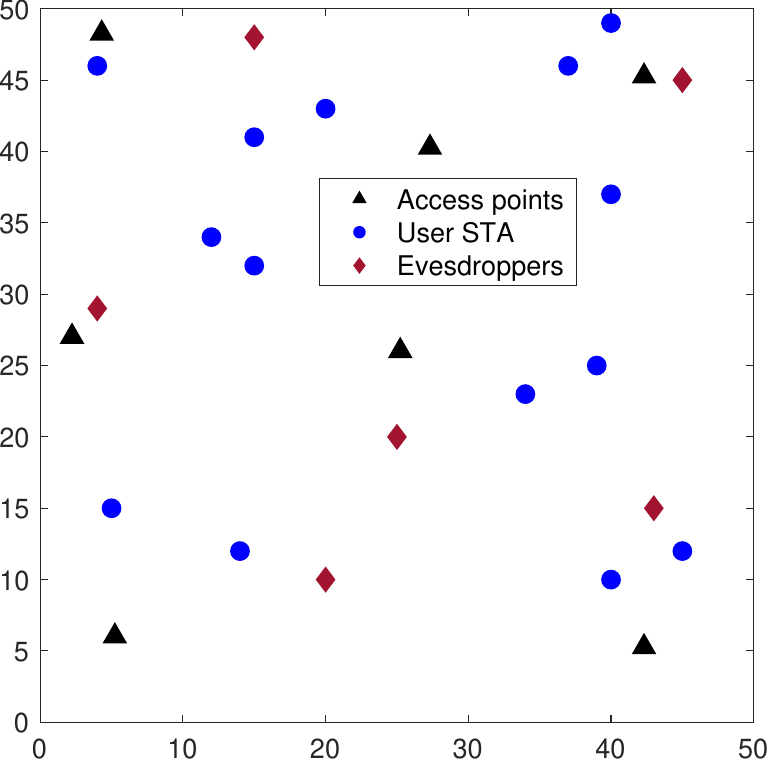}
        \caption{scenario 4: 7 APs.}
    \end{subfigure} \hspace{25pt}
    \begin{subfigure}{0.27\textwidth}
        \includegraphics[trim=0 0 0 0, clip, height=5.6cm]{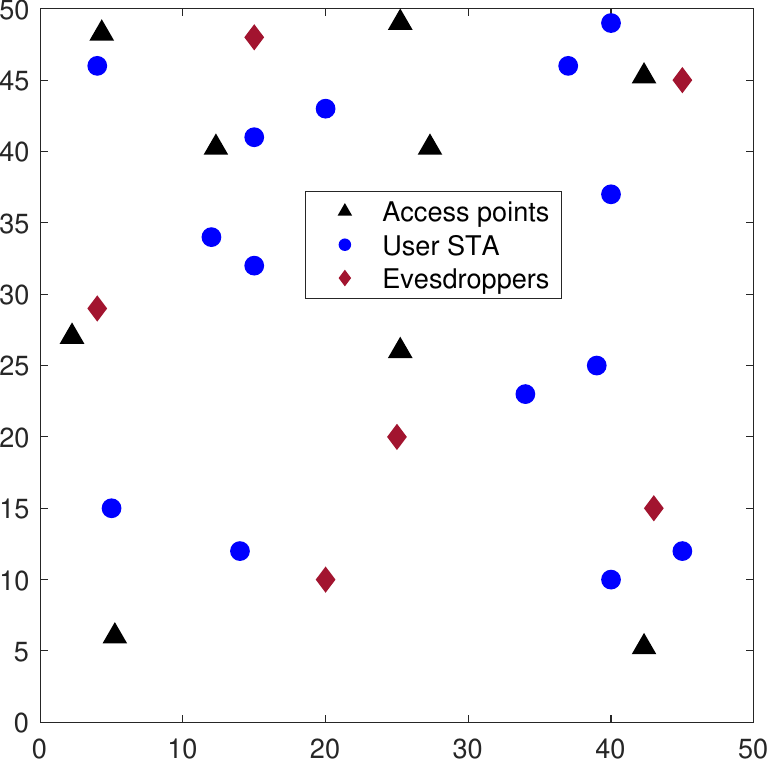}
        \caption{scenario 5: 9 APs.}
    \end{subfigure} \hspace{25pt}
    \begin{subfigure}{0.27\textwidth}
        \includegraphics[trim=0 0 0 0, clip, height=5.6cm]{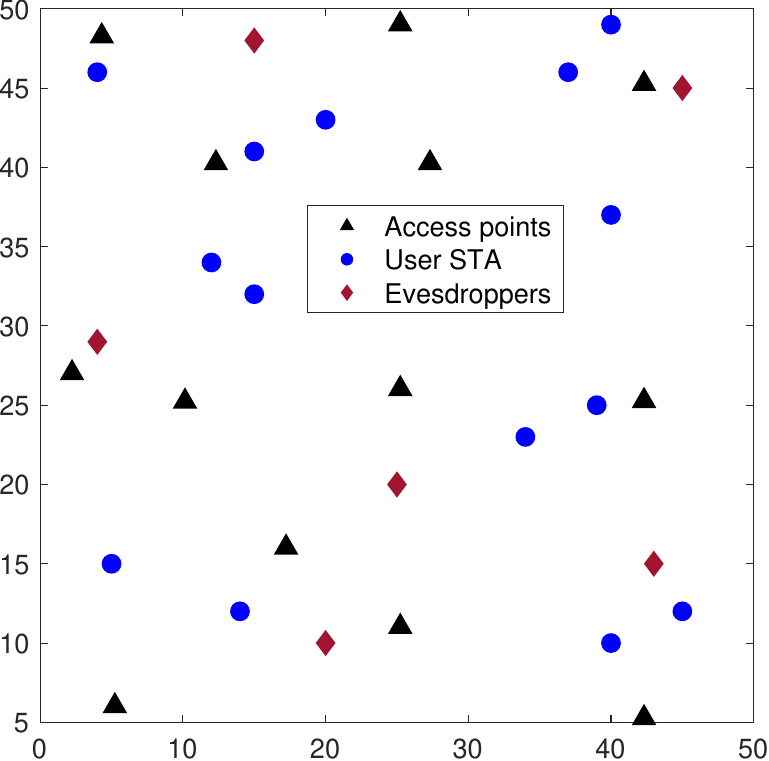}
        \caption{scenario 6: 13 APs.}
    \end{subfigure}  
    % \begin{subfigure}{0.4\textwidth}
    %     \includegraphics[trim=0 0 0 0, clip, height=5.6cm]{figures/sch_sc8.pdf}
    %     \caption{scenario 8}
    % \end{subfigure}  
    \caption{These figures display the locations of users, eavesdroppers, and APs in various scenarios. Scenarios 3, 4, 5, and 6 involve identical numbers and locations of users and eavesdroppers, with only new APs being introduced.}
    \label{fig:scenarios}
\end{figure*}

 We evaluated the performance of the proposed RL-based CFJ  model using simulations with Matlab. We simulated three Wi-Fi systems: a) a normal Wi-Fi implementation where the user is associated with the AP with the highest SINR regardless of the eavesdroppers' location and eavesdropping capacity; b) the smart AP implementation where the user is associated with the AP that provides the highest secrecy capacity according to \eqref{eq:AP_sel} (but without any power optimization); and c) The RL-Based CFJ implementation where each AP can work as a source of legitimate traffic or jamming to increase the secrecy of communication according to \eqref{eq:AP_sel} and \eqref{eq:sum_cam_opt_power} where its transmit power is optimized. In more detail, each AP is sending either data frames or jamming signals continuously. If APs associate with more than one user, time division multiplexing will be used. However, instantaneous capacity is considered for secrecy. As shown in~\eqref{eq:eve_cap}, each AP can contribute to the friendly jamming effectiveness against eavesdropping on other AP's traffic. In the normal Wi-Fi (a) and smart AP (b) implementation, the transmit power of every AP is set to 1 watt, and the idle APs (not associated with any legitimate user) are neither transmitting data traffic nor jamming. The RL-based (c) implementation firstly employs \eqref{eq:AP_sel} to select the AP for each user assuming the transmit powers are maximum across APs and then optimizes transmit powers to achieve the maximum sum secrecy capacity. The optimal transmit power is limited to a maximum of 1 Watt.

\begin{figure*}
   \centering
    \begin{subfigure}{0.28\textwidth}
        \includegraphics[trim=0 0 0 0, clip, height=4.8cm]{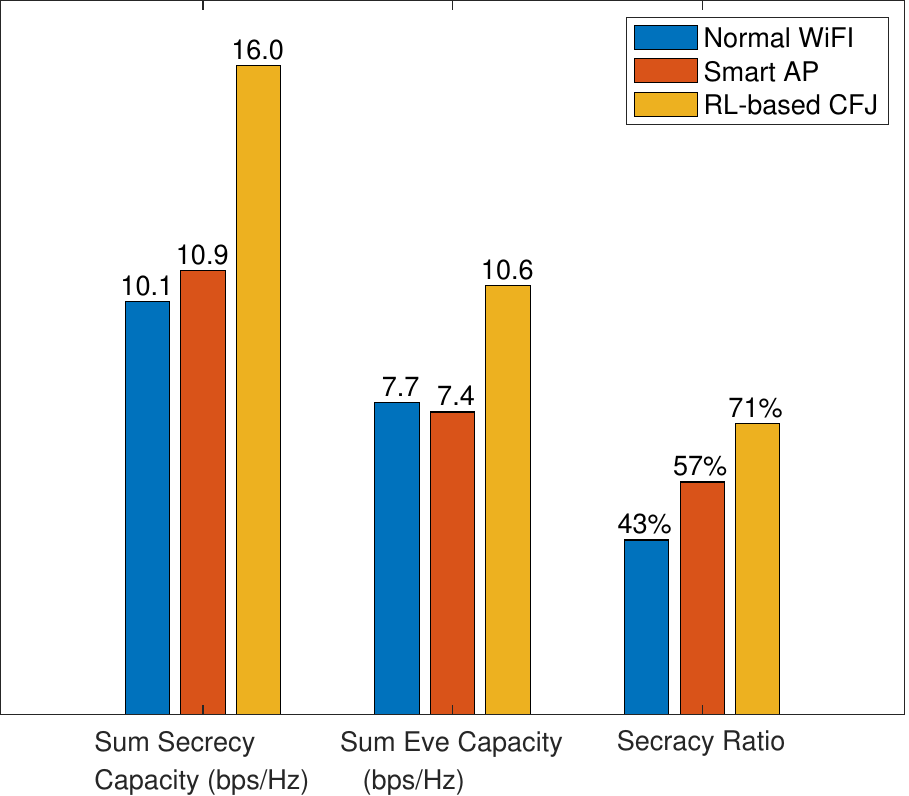}
        \caption{scenario 1} 
    \end{subfigure} \hspace{0.5cm}
        \begin{subfigure}{0.28\textwidth}
        \includegraphics[trim=0 0 0 0, clip, height=4.8cm]{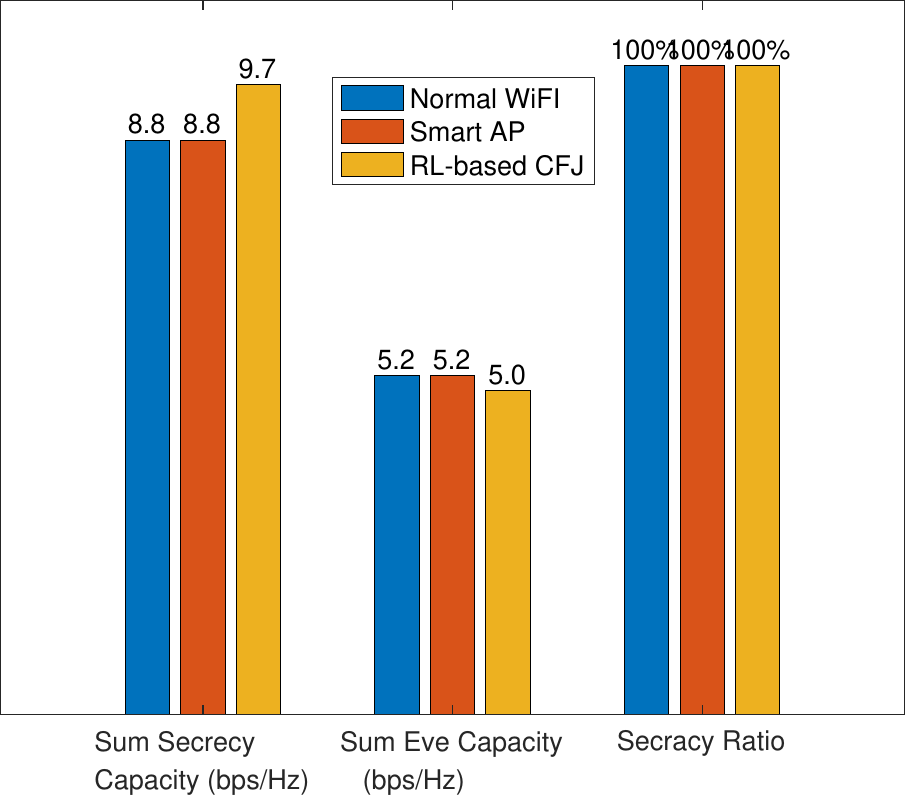}
        \caption{scenario 2}
    \end{subfigure} \hspace{0.5cm}
    % \begin{subfigure}{0.4\textwidth}
    %     \includegraphics[trim=0 0 0 0, clip, height=5.6cm]{figures/re_sc3.pdf}
    %     \caption{scenario 3}
    % \end{subfigure}
    \begin{subfigure}{0.28\textwidth}
        \includegraphics[trim=0 0 0 0, clip, height=4.8cm]{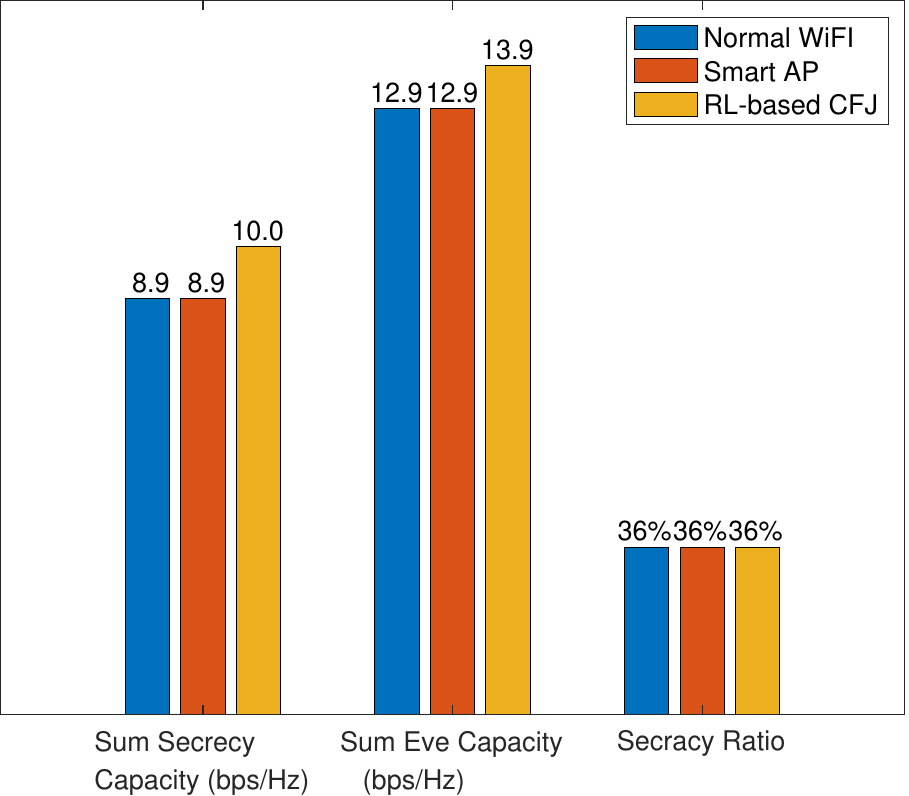}
        \caption{scenario 3}
    \end{subfigure} \\
    \begin{subfigure}{0.28\textwidth}
        \includegraphics[trim=0 0 0 0, clip, height=4.8cm]{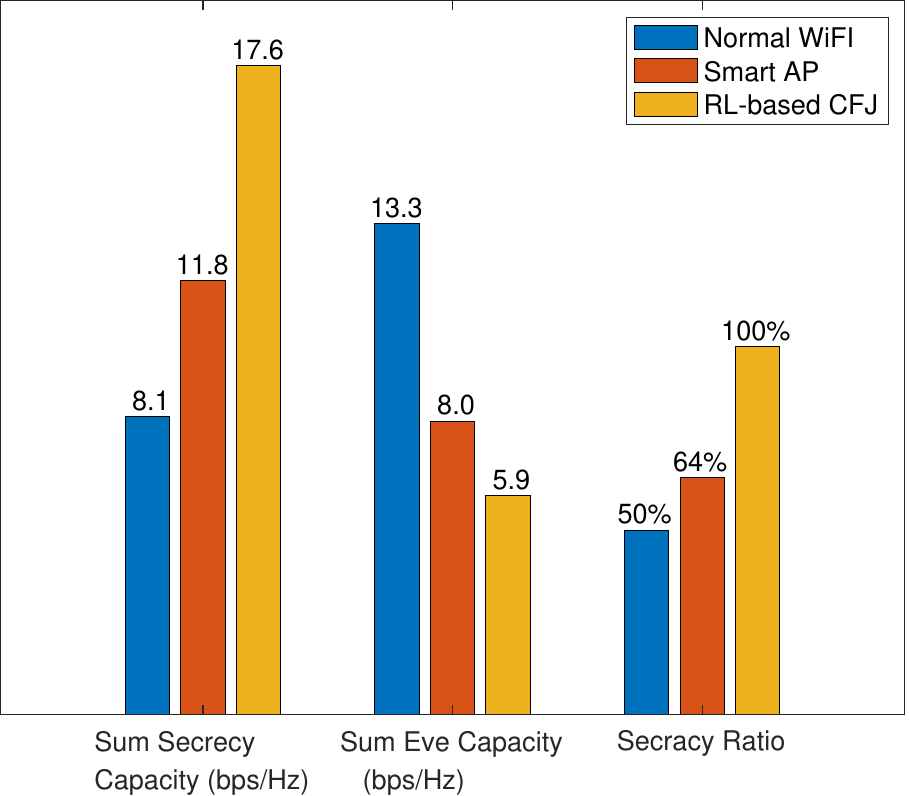}
        \caption{scenario 4}
    \end{subfigure}   \hspace{0.5cm}
    \begin{subfigure}{0.28\textwidth}
        \includegraphics[trim=0 0 0 0, clip, height=4.8cm]{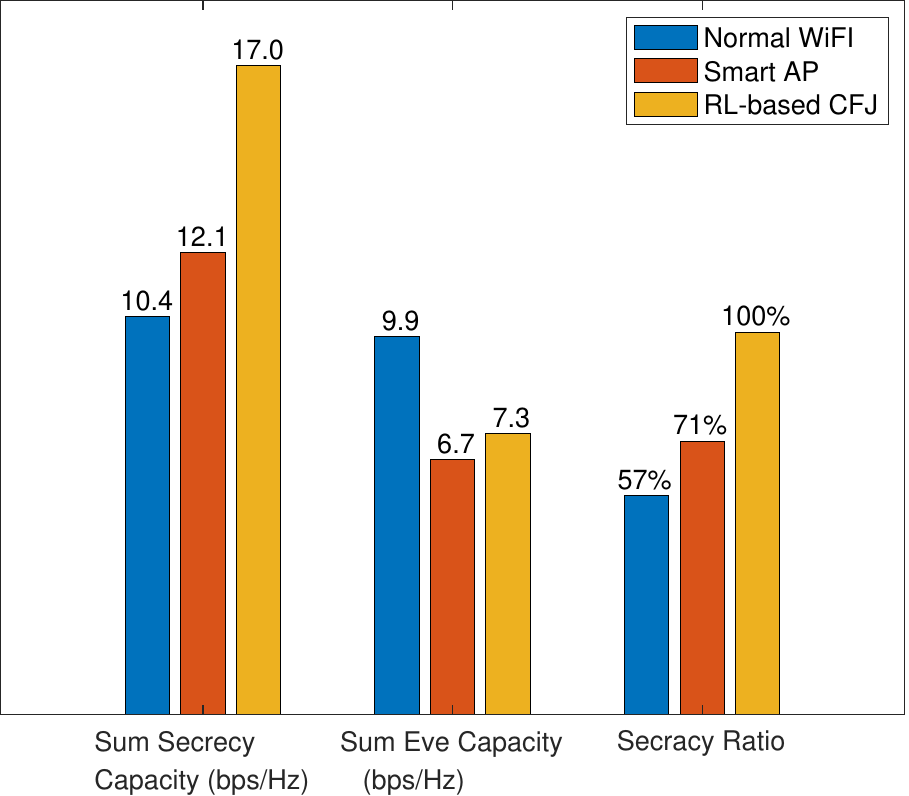}
        \caption{scenario 5}
    \end{subfigure} \hspace{0.5cm}
    \begin{subfigure}{0.28\textwidth}
        \includegraphics[trim=0 0 0 0, clip, height=4.8cm]{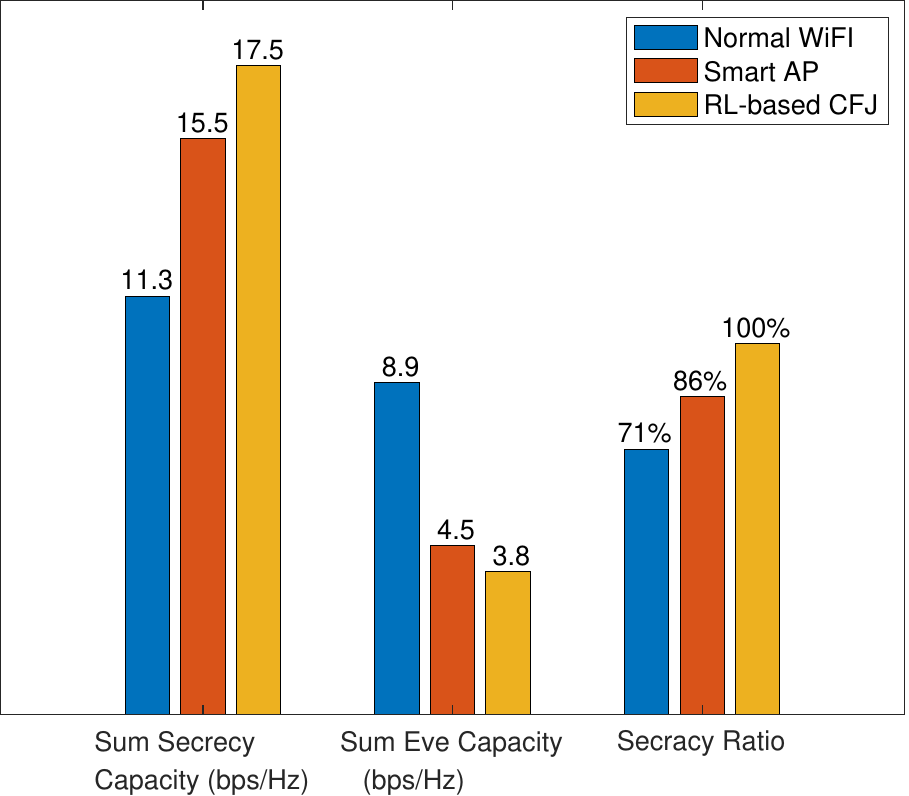}
        \caption{scenario 6}
    \end{subfigure} 
    % \begin{subfigure}{0.4\textwidth}
    %     \includegraphics[trim=0 0 0 0, clip, height=5.6cm]{figures/re_sc8.pdf}
    %     \caption{scenario 8}
    % \end{subfigure} 
    \caption{Results for different scenarios for 3 different implementations. Rl-based CFJ can generally outperform other implementations. Scenarios 3, 4, 5, and 6 involve identical numbers and locations of users and eavesdroppers, with only new APs being introduced. Therefore, secrecy can be improved by deploying more APs. Although scenario 5 is expected to perform as well as or better than scenario 4 for RL-based CFJ, the results show a slight downgrade. This is because the AP selection algorithm is run separately from power optimization and before it.}
    \label{fig:results}
\end{figure*}

 The map considered for the simulations has dimensions of $50m \times 50m$, where all coordinates are expressed in meters (m). Users and eavesdroppers were randomly located, and the APs were placed sufficiently far from each other to cover the entire map area. The communication between users and APs is conducted over Wi-Fi operating at $f_0 = 2.4$ GHz. The noise power at all nodes was set to $N = N_m = N_e = -85$ dBm $= 3.16\times10^{-12}$ Watts. We assumed a path-loss exponent of $\gamma = 2$ for the entire map. To simulate the environment and train the RL agent, we used Matlab's Reinforcement Learning Toolbox. The action, as defined in Section~\ref{sec:RL-SAC}, was implemented as a continuous space. The map was divided into $1m \times 1m$ cells to discretize the observation space and limit possible states. The deep learning networks for the SAC agent's critic and actors comprised nine layers of depth and 32 to 256 hidden units, depending on the number of APs.

In order to investigate the impact of the user node and AP location on our system, we conducted simulations of six different scenarios, as presented in \figurename~\ref{fig:scenarios}. Scenarios 3 to 6 featured the same number of users and eavesdroppers, all located in the same positions. However, the number of APs varied, increasing from 5 in scenario 3 to 7, 9, and 13 in scenarios 4, 5, and 6, respectively. This allowed us to explore the effects of more densely deployed APs on our system. For each scenario, we simulated the three implementations of normal Wi-Fi, smart AP and RL-based CFJ  as described above. For the results, we calculated a) \emph{sum secrecy capacity} $\sum^{K}_{k=1} \max(C^{s}(u_k,\alpha_k)$, b) \emph{sum Eve capacity} where the capacity of each eavesdropper is the highest eavesdropping capacity on $K$ users according to \eqref{eq:maxEveCap} and c) the \emph{secrecy ratio}, which we define as the percentage of users that can achieve positive secrecy capacity. The capacities are shown for $W=1$ Hz or equivalently bit per second per hertz (bps/Hz). The antenna gains at the transmitter, $G_t$, and the receiver, $G_r$, are assumed to be unity. 

 The results are visualized as in \figurename~\ref{fig:results}. It can be observed that the RL-based CFJ outperforms the normal Wi-Fi and smart AP implementations in all scenarios, achieving a higher sum secrecy capacity. Please note that the revenue function in \eqref{eq:revenue} only rewards the sum secrecy capacity according to \eqref{eq:sum_cam_opt} and there is no explicit penalty for eavesdropper's capacity in this equation. Nevertheless, such RL method also manages to achieve a lower eavesdropping capacity for most scenarios. To demonstrate the distribution of secrecy capacity improvement among users, we presented another group of bar charts depicting the percentage of secrecy ratio in Figure \ref{fig:results}. The results clearly show that the proposed RL-based model outperforms the other two implementations. Moreover, the smart AP implementation also shows better performance than normal Wi-Fi as the latter does not take into account the PLS in any form.

The second scenario results are distinct in the sense that we can see for all the implementations, the secrecy ratio is 100\%. This is due to the eavesdroppers' location, where almost all users are closer than the eavesdroppers to at least one AP. This leads to positive secrecy for all users in all implementations where the smart AP selection as presented in \eqref{eq:AP_sel} results in the same selection as normal Wi-Fi (where the closest AP is simply selected). 

In the third scenario, there is an increase in both the number of users and eavesdroppers, leading to a significant increase in eavesdropping capacity and a decrease in secrecy ratio. RL-Based power optimization also cannot improve the secrecy capacity notably. However, by adding two more APs to scenario 3 and keeping the same number and location for users and eavesdroppers, it can be observed in scenario 4 that there is a significant improvement in the results. The proposed RL-based implementation remarkably outperforms the other two approaches.

However, by adding two more APs in scenario 5, the sum secrecy capacity is slightly lower and the eavesdropper capacity increases for  RL-based results. This was unexpected, given that we anticipated achieving at least the same level of performance as scenario 4. Our investigation revealed this is because some users are now associated with newly added APs. We conclude that since the AP selection algorithm is done separately and first, by assuming uniform power allocation, and then the RL model is used to find the optimal power vector, the whole process is not optimal. This should be improved by integrating the AP selection in the RL model in future work. Generally speaking, we can conclude that having a more dense AP deployment improves the secrecy of users as can be seen in scenarios 3 to 6.

\section{Conclusions and Future Work}
\label{sec:conc}
In this paper, we presented an optimization solution for the transmit power allocation, necessary to achieve CFJ in large Wi-Fi networks, by leveraging the Soft Actor-Critic RL agent methodology within reinforcement learning. The obtained results show that the proposed solution enhances the secrecy of wireless communication in the presence of multiple eavesdroppers. Furthermore, we found that a higher density of APs generally offers more opportunities to improve the secrecy of communication.

These results demonstrate the effectiveness of ML in addressing the complex problem of transmit power optimization of friendly jamming to achieve PLS. They also showcase the potential of ML in addressing other optimization problems in PLS that could not be solved with conventional theoretical or numerical optimization methods. More specifically, we aim to integrate the AP selection algorithm into the RL model in our future work, to further optimize the secrecy of the wireless network. Finally, we are also considering the extension of this work to mobile legitimate stations and eavesdroppers, by training the RL model to the dynamic environment and mobility patterns that characterize large Wi-Fi networks. 

\bibliographystyle{IEEEtran}
\bibliography{References}

\begin{thebibliography}{10}
\providecommand{\url}[1]{#1}
\csname url@samestyle\endcsname
\providecommand{\newblock}{\relax}
\providecommand{\bibinfo}[2]{#2}
\providecommand{\BIBentrySTDinterwordspacing}{\spaceskip=0pt\relax}
\providecommand{\BIBentryALTinterwordstretchfactor}{4}
\providecommand{\BIBentryALTinterwordspacing}{\spaceskip=\fontdimen2\font plus
\BIBentryALTinterwordstretchfactor\fontdimen3\font minus
  \fontdimen4\font\relax}
\providecommand{\BIBforeignlanguage}[2]{{%
\expandafter\ifx\csname l@#1\endcsname\relax
\typeout{** WARNING: IEEEtran.bst: No hyphenation pattern has been}%
\typeout{** loaded for the language `#1'. Using the pattern for}%
\typeout{** the default language instead.}%
\else
\language=\csname l@#1\endcsname
\fi
#2}}
\providecommand{\BIBdecl}{\relax}
\BIBdecl

\bibitem{liu2016physical}
Y.~Liu, H.-H. Chen \emph{et~al.}, ``Physical layer security for next generation
  wireless networks: Theories, technologies, and challenges,'' \emph{IEEE
  Communications Surveys \& Tutorials}, vol.~19, no.~1, pp. 347--376, 2016.

\bibitem{hoseini2022ccnc}
S.~A. Hoseini, F.~Bouhafs \emph{et~al.}, ``A practical implementation of
  physical layer security in wireless networks,'' in \emph{2022 IEEE 19th
  Annual Consumer Communications Networking Conference (CCNC)}, 2022, pp. 1--4.

\bibitem{hoseini2023demo}
------, ``{Realizing Physical Layer Security with common off-the-shelf WiFi
  equipment},'' in \emph{CCNC 2023 : IEEE Consumer Communications and
  Networking Conference}, 2023.

\bibitem{bouhafs2018wi}
F.~Bouhafs, M.~Mackay \emph{et~al.}, ``Wi-5: A programming architecture for
  unlicensed frequency bands,'' \emph{IEEE Communications Magazine}, vol.~56,
  no.~12, pp. 178--185, 2018.

\bibitem{Faycal2020globecom}
F.~Bouhafs, F.~den Hartog \emph{et~al.}, ``Realizing physical layer security in
  large wireless networks using spectrum programmability,'' in \emph{2020 IEEE
  Globecom Workshops}, 2020, pp. 1--6.

\bibitem{hoseini2022jamming}
S.~A. Hoseini, P.~Sadeghi \emph{et~al.}, ``{Network-Controlled Physical-Layer
  Security: Enhancing Secrecy Through Friendly Jamming},'' in \emph{27th IEEE
  Symposium on Computers and Communications (ISCC 2022)}, 2022.

\bibitem{RN1}
T.~Erpek, T.~O’Shea \emph{et~al.}, ``Deep learning for wireless
  communications,'' \emph{Development and Analysis of Deep Learning
  Architectures}, pp. 223--266, 2020.

\bibitem{RN2}
D.~Janu, K.~Singh \emph{et~al.}, ``Machine learning for cooperative spectrum
  sensing and sharing: A survey,'' \emph{Transactions on Emerging
  Telecommunications Technologies}, vol.~33, no.~1, p. e4352, 2022.

\bibitem{RN3}
M.~Ahmed, B.~Gabr \emph{et~al.}, ``Machine learning-based module for monitoring
  lte/wifi coexistence networks dynamics,'' in \emph{2021 IEEE International
  Conference on Communications Workshops (ICC Workshops)}.\hskip 1em plus 0.5em
  minus 0.4em\relax IEEE, Conference Proceedings, pp. 1--6.

\bibitem{RN4}
K.~Youssef, L.~Bouchard \emph{et~al.}, ``Machine learning approach to rf
  transmitter identification,'' \emph{IEEE Journal of Radio Frequency
  Identification}, vol.~2, no.~4, pp. 197--205, 2018.

\bibitem{RN5}
Y.~Arjoune, F.~Salahdine \emph{et~al.}, ``A novel jamming attacks detection
  approach based on machine learning for wireless communication,'' in
  \emph{2020 International Conference on Information Networking (ICOIN)}.\hskip
  1em plus 0.5em minus 0.4em\relax IEEE, Conference Proceedings, pp. 459--464.

\bibitem{RN6}
G.~Kasturi, A.~Jain \emph{et~al.}, ``Detection and classification of radio
  frequency jamming attacks using machine learning,'' \emph{J. Wirel. Mob.
  Networks Ubiquitous Comput. Dependable Appl.}, vol.~11, no.~4, pp. 49--62,
  2020.

\bibitem{RN7}
O.~Puñal, I.~Aktaş \emph{et~al.}, ``Machine learning-based jamming detection
  for ieee 802.11: Design and experimental evaluation,'' in \emph{Proceeding of
  IEEE International Symposium on a World of Wireless, Mobile and Multimedia
  Networks 2014}.\hskip 1em plus 0.5em minus 0.4em\relax IEEE, Conference
  Proceedings, pp. 1--10.

\bibitem{RN8}
A.~K. Kamboj, P.~Jindal \emph{et~al.}, ``Machine learning-based physical layer
  security: techniques, open challenges, and applications,'' \emph{Wireless
  Networks}, vol.~27, pp. 5351--5383, 2021.

\bibitem{RN9}
L.~Zhang, S.~Lai \emph{et~al.}, ``Deep reinforcement learning based
  irs-assisted mobile edge computing under physical-layer security,''
  \emph{Physical Communication}, vol.~55, p. 101896, 2022.

\bibitem{wang2022secrecy}
T.~Wang, Y.~Li \emph{et~al.}, ``Secrecy driven federated learning via
  cooperative jamming: An approach of latency minimization,'' \emph{IEEE
  Transactions on Emerging Topics in Computing}, vol.~10, no.~4, pp.
  1687--1703, 2022.

\bibitem{haarnoja2018SAC}
T.~Haarnoja, A.~Zhou \emph{et~al.}, ``Soft actor-critic: Off-policy maximum
  entropy deep reinforcement learning with a stochastic actor,'' in
  \emph{International conference on machine learning}.\hskip 1em plus 0.5em
  minus 0.4em\relax PMLR, 2018, pp. 1861--1870.

\bibitem{eletreby2015supporting}
R.~Eletreby, H.~Rahbari \emph{et~al.}, ``Supporting phy-layer security in
  multi-link wireless networks using friendly jamming,'' in \emph{2015 IEEE
  Global Communications Conference (GLOBECOM)}.\hskip 1em plus 0.5em minus
  0.4em\relax IEEE, 2015, pp. 1--6.

\bibitem{adams2017FJ}
M.~Adams and V.~K. Bhargava, ``Using friendly jamming to improve route security
  and quality in ad hoc networks,'' in \emph{2017 IEEE 30th Canadian Conference
  on Electrical and Computer Engineering (CCECE)}, 2017, pp. 1--6.

\bibitem{rappaport2002mobile}
T.~S. Rappaport, ``Mobile radio propagation: Large-scale path loss,''
  \emph{Wireless communications: principles and practice}, pp. 107--110, 2002.

\bibitem{cover2006elements}
T.~M. Cover, \emph{Elements of information theory}.\hskip 1em plus 0.5em minus
  0.4em\relax John Wiley \& Sons, 2006.

\bibitem{wyner1975wire}
A.~D. Wyner, ``The wire-tap channel,'' \emph{Bell system technical journal},
  vol.~54, no.~8, pp. 1355--1387, 1975.

\bibitem{vilela2011wireless}
J.~P. Vilela, M.~Bloch \emph{et~al.}, ``Wireless secrecy regions with friendly
  jamming,'' \emph{IEEE Transactions on Information Forensics and Security},
  vol.~6, no.~2, pp. 256--266, 2011.

\bibitem{PLS_VLC_2022}
F.~Yang, J.~Wang \emph{et~al.}, ``Physical-layer security for indoor {VLC}
  wiretap systems under multipath reflections,'' \emph{IEEE Transactions on
  Wireless Communications}, vol.~21, no.~12, pp. 11\,179--11\,192, 2022.

\bibitem{zhang2013optimization}
M.~Zhang, R.~Xue \emph{et~al.}, ``Secrecy capacity optimization in coordinated
  multi-point processing,'' in \emph{2013 IEEE International Conference on
  Communications (ICC)}, 2013, pp. 5845--5849.

\bibitem{SACopenAI}
\BIBentryALTinterwordspacing
``Soft actor-critic (sac).'' [Online]. Available:
  \url{https://spinningup.openai.com/en/latest/algorithms/sac.html}
\BIBentrySTDinterwordspacing

\end{thebibliography}
\end{document}